\documentclass{bmcart}

\usepackage[utf8]{inputenc} 
\usepackage{graphicx}
\usepackage{graphics}

\startlocaldefs
\endlocaldefs

\begin{document}

\begin{frontmatter}

\begin{fmbox}
\dochead{Research}


\title{Backbone of credit relationships in the Japanese credit market}


\author[
   addressref={aff1},                   
   email={marotta.luca@gmail.com}   
]{\inits{L}\fnm{Luca} \snm{Marotta}}
\author[
   addressref={aff1},
   email={salvatore.micciche@unipa.it}
]{\inits{S}\fnm{Salvatore} \snm{Miccich\'e}}
\author[
   addressref={aff2},
   email={yoshi.fujiwara@gmail.com}
]{\inits{Y}\fnm{Yoshi} \snm{Fujiwara}}
\author[
   addressref={aff3},
   email={hiyetomi@gmail.com}
]{\inits{H}\fnm{Hiroshi} \snm{Iyetomi}}
\author[
   addressref={aff4},
   email={hideaki.aoyama@gmail.com}
]{\inits{H}\fnm{Hideaki} \snm{Aoyama}}
\author[
   addressref={aff5},
   email={mauro.gallegati@gmail.com}
]{\inits{M}\fnm{Mauro} \snm{Gallegati}}
\author[
   addressref={aff1,aff6},
   corref={aff1},                       
   email={rn.mantegna@gmail.com}   
]{\inits{RN}\fnm{Rosario N.} \snm{Mantegna}}

\address[id=aff1]{
  \orgname{Dipartimento di Fisica e Chimica, Universit\'a degli Studi di Palermo}, 
  \street{Viale delle Scienze, Ed. 18},                     %
  \postcode{I-90128}                                
  \city{Palermo},                              
  \cny{Italia}                                    
}
\address[id=aff2]{%
  \orgname{Graduate School of Simulation Studies, The University of Hyogo},
  \postcode{650-0047}
  \city{Kobe},
  \cny{Japan}
}
\address[id=aff3]{%
  \orgname{Department of Mathematics, Niigata University},
  \postcode{950-2181}
  \city{Niigata},
  \cny{Japan}
}
\address[id=aff4]{%
  \orgname{Graduate School of Sciences, Kyoto University},
  \postcode{606-8502}
  \city{Kyoto},
  \cny{Germany}
}
\address[id=aff5]{%
  \orgname{Dipartimento di Scienze Economiche e Sociali, Universit\`a Politecnica delle Marche},
  \street{Piazza Martelli, 8},
  \postcode{60121}
  \city{Ancona},
  \cny{Italy}
}
\address[id=aff6]{%
  \orgname{Center for Network Science and Department of Economics, Central European University},
  \street{Nador 9},
  \postcode{1051}
  \city{Budapest},
  \cny{Hungary}
}


\begin{artnotes}
\end{artnotes}

\end{fmbox}


\begin{abstractbox}

\begin{abstract} 
We detect the backbone of the weighted bipartite network of the Japanese credit market  relationships. The backbone is detected by adapting a general method used in the investigation of weighted networks. With this approach we detect a backbone that is statistically validated against a null hypothesis of uniform diversification of loans for banks and firms. Our investigation is done year by year and it covers more than thirty years during the period from 1980 to 2011. We relate some of our findings with economic events that have characterized the Japanese credit market during the last years. The study of the time evolution of the backbone allows us to detect changes occurred in network size, fraction of credit explained, and attributes characterizing the banks and the firms present in the backbone. 
\end{abstract}


\begin{keyword}
\kwd{Complex networks}
\kwd{Information filtering}
\kwd{Statistically validated networks}
\kwd{Credit market}
\end{keyword}


\end{abstractbox}

\end{frontmatter}




\section*{Introduction}
Bipartite networks are observed in a wide variety of disciplines. In economic complex systems we find, for instance, 
the bipartite networks of boards and directors  \cite{Battiston2003,Battiston2004} and products and consumers 
\cite{Ueno2011,Pennacchioli2014}. 
In this paper we investigate the bipartite network of credit relationships  between all Japanese banks and large firms listed at 
Japanese stock exchanges and over-the-counter markets of Japan. 

This credit system has been previously analyzed with tools and concepts of network theory in a series of studies.  Examples are the study of the one-mode projected network \cite{DeMasi}, the study of the eigenvalue problem determined by the weight of the credit network \cite{Fujiwara2009}, the use of the debt-rank concept to analyze the risk and fragility of the credit market \cite{Aoyama2013} 
and the study of the communities observed in the projected and  bipartite networks \cite{Iyetomi2013, Marotta2015}. 

Here, we inspect the weighted bipartite network by filtering the credit relationships that are not compatible with a null hypothesis assuming uniform diversification of each bank or firm. The method we use is adapted from the general method proposed in ref. \cite{Serrano2009}. We add to the procedure proposed in ref. \cite{Serrano2009} a multiple hypothesis correction that we believe is necessary to minimize false positive \cite{Tumminello2011b}. With this approach we detect networks that are statistically validated against the chosen null hypothesis.

Our investigation covers more than thirty years giving us the possibility to relate some of our findings with documented events such as major failures of the credit system and big mergers that occurred in the Japanese credit market. All the statistically validated networks are then characterized in terms of over-expression of attributes of banks and firms concerning (i) the types of banks,  (ii) economic sectors of firms, and the geographical location of firms. The over-expression of such attributes is estimated according to the method introduced in ref. \cite{Tumminello2011}.

We detect the presence of a backbone of the credit relationships which is not compatible with the null hypothesis of uniform diversification. This backbone is present for all the investigated years. During the different years we are able to see the changes occurred in this filtered network both in size, fraction of credit explained and attributes characterizing the firms present in it. 

The paper is organized as follows. In Sect. ``The dataset'' we describe our dataset. The Sect. ``The filtering methodology'' briefly discusses the filtering method and the need for a multiple hypothesis correction. In the successive Sect. ``Results'' we present and comment our results. In Sect. ``Conclusions'' we briefly draw some conclusions.

\section*{The dataset}
Our dataset was obtained by Nikkei Media Marketing, Inc. in Tokyo, and are commercially available (see the webpage \cite{Nikkei} for details). Data are based on a survey of firms quoted in the Japanese stock-exchanges  (Tokyo, Osaka, Nagoya) and in Japanese over-the-counter (OTC) markets . Data includes information about each firm's borrowing from banks obtained from financial institutions. Specifically, the dataset reports the amounts of borrowing of each firm and the classification of loans into short-term and long-term borrowings. All contracts 
exceeding 1 year are considered long-term borrowing. Data covers the time period from 1980 to 2012. In this paper we examine the time period from 1980 to 2011, which is a time period of more than three decades. Our analyses are performed yearly, and yearly networks are constructed from the dataset by using the financial statements of the considered calendar year. Since 1996 the dataset includes also firms listed at the OTC markets and/or at the JASDAQ (the present OTC market).  In the present study we investigate all firms which are present in the database.

The number of banks of the database changes year by year. It was 225 in 1980, remained approximately constant until 2001 and then decreased to 166 in 2011. The number of firms was starting from the value of 1414 in 1980 and then increasing to the maximal value of 3034 reached in 2006. After this year the number of firms started to decrease and reached the value of 2706 in 2011. The number of firms increased from 1802 in 1995 to 2602 in 1996 in the presence of the largest inclusion of the OTC firms in the database. During the same years the number of banks increased from 219 to 226. The density of links in the bipartite network, defined as number of observed links over number of potential links, was on average decreasing from the value of 0.0867 in 1980 to the value of 0.0398 in 2011. The variation of the density of links was not too large over the years including the years of the largest inclusion of the OTC firm. In fact the density of links decreased from 0.0721 to 0.0601 from 1995 to 1996.   

The dataset has metadata associated. Specifically we have information about the classification of each bank and information about the economic sector and geographical location of firms. 

\section*{The filtering methodology}
In this paper we wish to focus on the credit relationships that are most relevant in terms of money allocated to the specific credit relationships for each node (bank or firm) of the bipartite network. A number of techniques have been proposed in order to single 
out the most important connections in a weighted network, such as (i) the application of a global threshold that would 
maintain the links whose weight is highest (though spoiling the intrinsic multi-scale organization of many complex systems)
\cite{Equiluz2005}, (ii) a method based on the statistical validation of a null hypothesis tested for each node  of the network \cite{Serrano2009}, and (iii) a method using a global null model preserving both the network topology and the weight distribution 
of the system \cite{Radicchi2011}.
   
In our present study, we are interested in the local anomalous distribution of the credit amount and not in preserving 
the network topology. Therefore we adapt the filtering procedure for weighted networks introduced in \cite{Serrano2009} to our system. 
This method assesses the relevance of the weight of each link by means of a statistical validation at the level of the single node. 
Given the local nature of the test, the method allows to preserve the heterogeneity of the weight distribution, 
thus overcoming the drawbacks of a global thresholding procedure. 

More in detail, let $s_i$ and $k_i$ be respectively the strength and the degree of node $i$, and let $w_{ij}$ be the weight of the link
between node $i$ and $j$. Denoting with $x_{ij}$ the normalized weight $w_{ij}/s_i$ of the link, the statistical procedure proposed 
in \cite{Serrano2009} answers the question: if we divide the interval [0, 1] in $k$ sub-intervals uniformly distributing $k-1$ 
points in it, what is the probability $p(x_{ij})$to observe an interval with length $x_{ij}$?

It can be shown that, under the above assumptions, the p-value is:
\begin{equation}
 \label{weight_prob}
 p=1-(k-1)\int_{0}^{x_{ij}}(1-x)^{k-2}dx.
\end{equation}

If $p$ in Eq. \ref{weight_prob} is smaller than a given, predetermined, statistical threshold $\theta$ the link is detected as statistically 
not consistent with the null hypothesis of a uniform distribution and therefore is preserved in the filtered network, otherwise it is 
deleted. Differently than in \cite{Serrano2009} we fix the statistical threshold and perform a multiple hypothesis test correction. 
In fact, due to the large number of tests needed to investigate the entire network a multiple hypothesis test correction is needed 
if one wants to minimize the number of false positive. In the present study we set our statistical threshold to the 
value $\theta=0.01$.

The most restrictive multiple hypothesis test correction is the Bonferroni correction.  \cite{Miller1981}, i.e. this correction is done by using as a statistical threshold $\theta_B=0.01/N_t$  instead of $\theta$, where $N_t$ is the number of test performed over the entire network. The Bonferroni correction increases the precision (by minimizing the number of false positive of the test) but decreases the accuracy 
of the estimation because can be associated with a large number of false negative. To avoid to be extremely restrictive, in this study we use the false discovery rate (FDR) \cite{Benjamini1995} as multiple hypothesis test correction.  The false discovery rate correction works as follows: the $p$-values of different tests are first arranged in increasing order ($p_1<p_2<...<p_t$) and then the FDR threshold is obtained 
by finding the largest $t_{max}$ such that $p_{t_{max}}<t_{max} ~ \theta_B$. It is worth stressing that, by construction, the Bonferroni 
network is always a subnetwork of the FDR network.     

Notably, the test in Eq. \ref{weight_prob} is directional, even for undirected networks. Indeed, 
the normalized weight $x_{ij}$ depends on the strength $s_i$ of node $i$ and therefore will be generally different 
from $x_{ji}$. This means each link has to be tested for both its end nodes, the presence of a validation in both directions signaling 
a strong inter-dependence between $i$ and $j$.

\begin{figure}[h!]
  \includegraphics[scale=0.4]{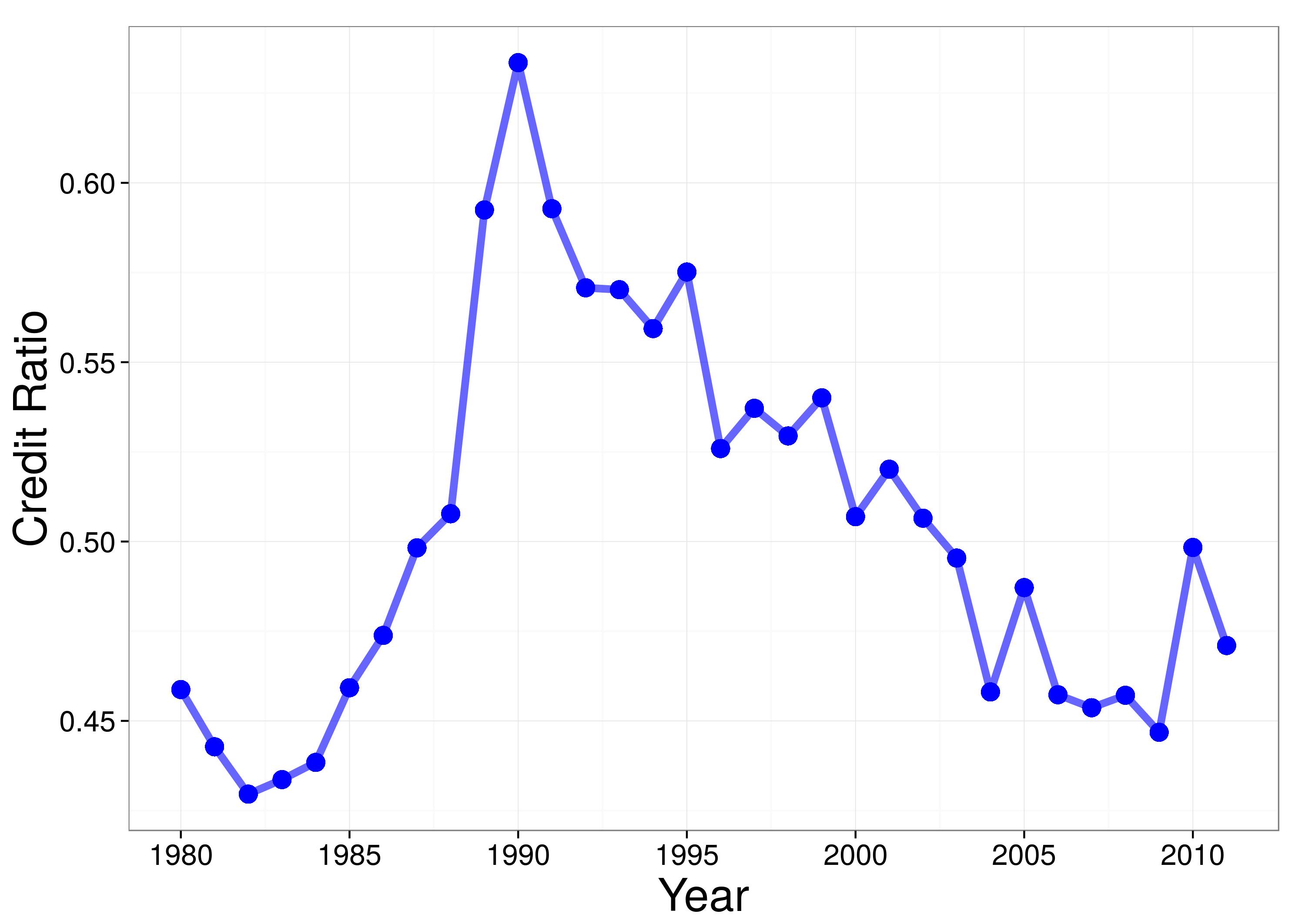}
  \caption{\csentence{Time evolution of the credit ratio}
     For each calendar year, the credit ratio is the ratio of the total amount of credit associated with the credit relationships selected in the statistically validated network divided by the total amount of credit exchanged in the system. It should be noted that the credit ration is increasing and reaching a maximum during the Japanese asset price bubble of 1986-1991.}
      \label{CreditRatio}
      \end{figure}

By using this approach we obtain for each year a FDR filtered bipartite network. A summary statistics with basic information about the original and filtered networks is given in Table \ref{Table1}. The Table shows that the filtering procedure is rather severe in fact the number of statistically validated credit relationships is on average of the order of 500 whereas the total number of credit relationships ranges from $35344$ in 1997 to $17885$ in 2011. In spite of that, the selected credit relationships are responsible for a fraction of the 
total credit lent by the banking system which is ranging from 45\% to more than 60\%. The time evolution of the credit ratio associated with the statistically validated edges is shown in Fig. \ref{CreditRatio}.  Quite interestingly, the credit ratio is growing from 1982 to 1990 and then is on average decreasing. It is probably worth noting that the 1990s was the year of the burst of the bubble of the Japanese stock markets. As a consequence of the collapse of the bubble the total bank assets declined from 508 bubble trillion yen in 1989 to about 491 trillion yen in 1990 \cite{Kanaya2000}.
For all the years, the statistically validated network present a largest connected component which is comprising a large percent of the elements of the bipartite network. Specifically this percent is ranging from the minimal value of 82\% observed in 2009 to the maximal value of 97\% observed in 1984. The number of elements included in the largest connected component of the statistically validated networks for each year are shown in the column LCC of Table \ref{Table1}. 

The statistical validation is performed for each node and therefore one credit link can be validated or with respect to a bank and/or with respect to a firms. We use the convention that a validated link is directed and the direction is outgoing from the node used in the validation procedure. For example if a credit relation is validated for a bank the arc will be outgoing from the bank and will point to the firm receiving the credit. The majority of the validated links are unidirectional but a certain fraction of bidirectional links, i.e. of credit relationships that are statistically validated both for the bank and for the firm, are also observed. Their number is shown in the column Pair BL of Table \ref{Table1} where we report the number of distinct pairs of bank-firm with bidirectional validated links. 
Some of these bidirectional validated links are long living, i.e. observed for as long as 21 years. These long living links are primarily observed during the period from 1980 to 2000. Their presence can be related to the existence of the so-called "main bank" relationships 
\cite{Kanaya2000,Brewer2003}. It should also be noted that the peak of the number of bank-firm pairs with credit relations relevant 
for both types of nodes is observed for 1997 which was the year of the full blown systemic crisis of the Japanese banking system \cite{Kanaya2000}      

\begin{table}[h!]
\caption{Summary statistics of the original bipartite networks (ON) and of the corresponding filtered networks (FN). Filtered networks are obtained by using the diversity filter of ref. \cite{Serrano2009} with $\theta=0.01$ and with the false discovery rate correction computed for each year.}
\label{Table1}
      \begin{tabular}{ccccccccc}
        \hline
          Year & Banks & Firms   & Edges & Banks& Firms& LCC & Edges & Pairs BL\\ 
 ~ & ON  & ON   & ON & FN & FN & FN & FN & FN\\           \hline
        1980 & 225 & 1414 & 27587 & 95 & 210 & 290 & 662 & 42\\
        1981 & 225 & 1431 & 27535 & 94 & 187 & 269 & 617 & 40\\
        1982 & 222 & 1444 & 27265 & 99 & 181 & 266 & 599 & 35\\
        1983 & 221 & 1457 & 26887 & 96 & 172 & 254 & 578 & 27\\
        1984 & 221 & 1462 & 26330 & 96 & 168 & 256 & 583 & 25\\
        1985 & 219 & 1477 & 25824 & 94 & 166 & 252 & 584 & 34\\
        1986 & 217 & 1486 & 25139 & 95 & 173 & 258 & 588 & 38\\
        1987 & 220 & 1530 & 25416 & 112 & 169 & 269 & 644 & 40\\
        1988 & 221 & 1545 & 25170 & 108 & 177 & 267 & 648 & 39\\
        1989 & 222 & 1573 & 25069 & 106 & 189 & 281 & 651 & 39\\
        1990 & 222 & 1617 & 25343 & 105 & 187 & 282 & 645 & 40\\
        1991 & 221 & 1670 & 25892 & 99 & 183 & 270 & 647 & 48\\
        1992 & 221 & 1687 & 26598 & 96 & 192 & 276 & 643 & 49\\
        1993 & 218 & 1717 & 27410 & 95 & 187 & 260 & 644 & 61\\
        1994 & 219 & 1753 & 27913 & 97 & 197 & 267 & 675 & 67\\
        1995 & 219 & 1802 & 28452 & 106 & 224 & 300 & 722 & 63\\
        1996 & 226 & 2602 & 35314 & 113 & 258 & 337 & 861 & 75\\
        1997 & 225 & 2726 & 35344 & 110 & 268 & 341 & 850 & 80\\
        1998 & 221 & 2772 & 35056 & 99 & 268 & 328 & 803 & 71\\
        1999 & 218 & 2869 & 35315 & 104 & 280 & 348 & 809 & 75\\
        2000 & 221 & 2942 & 29565 & 80 & 230 & 284 & 627 & 50\\
        2001 & 210 & 2975 & 28719 & 80 & 230 & 285 & 625 & 49\\
        2002 & 207 & 2991 & 26610 & 71 & 212 & 253 & 556 & 35\\
        2003 & 199 & 2963 & 24564 & 79 & 216 & 250 & 501 & 32\\
        2004 & 197 & 2959 & 23888 & 72 & 195 & 233 & 463 & 23\\
        2005 & 190 & 3003 & 23903 & 73 & 190 & 232 & 464 & 27\\
        2006 & 185 & 3034 & 23012 & 75 & 187 & 231 & 441 & 23\\
        2007 & 181 & 3016 & 22273 & 69 & 186 & 231 & 417 & 18\\
        2008 & 178 & 2918 & 20567 & 67 & 183 & 227 & 389 & 17\\
        2009 & 177 & 2842 & 19229 & 71 & 184 & 209 & 346 & 9\\
        2010 & 175 & 2746 & 18357 & 61 & 173 & 204 & 334 & 10\\
        2011 & 166 & 2706 & 17885 & 64 & 158 & 182 & 323 & 9\\
        \hline
      \end{tabular}
\end{table}

\section*{Results}
\begin{figure}[h!]
  {\includegraphics[scale=0.23]{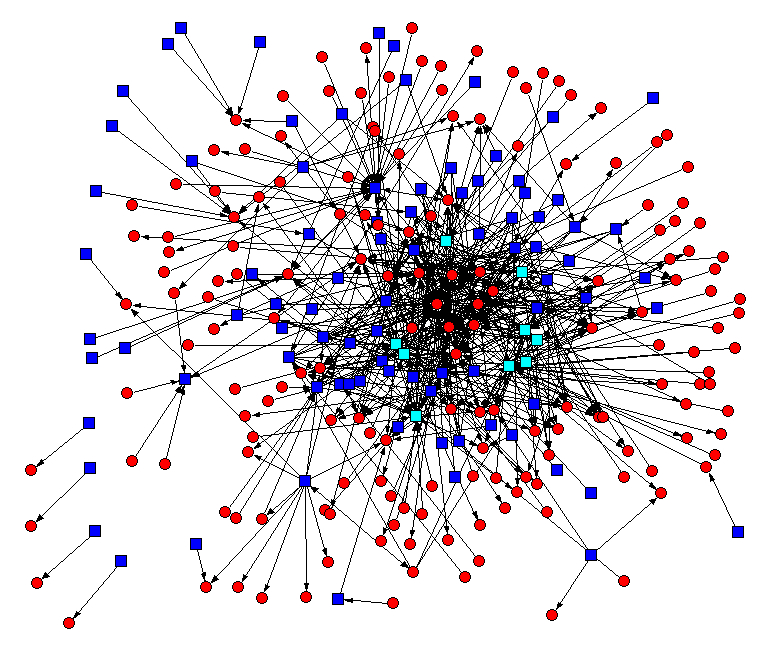}
  \includegraphics[scale=0.22]{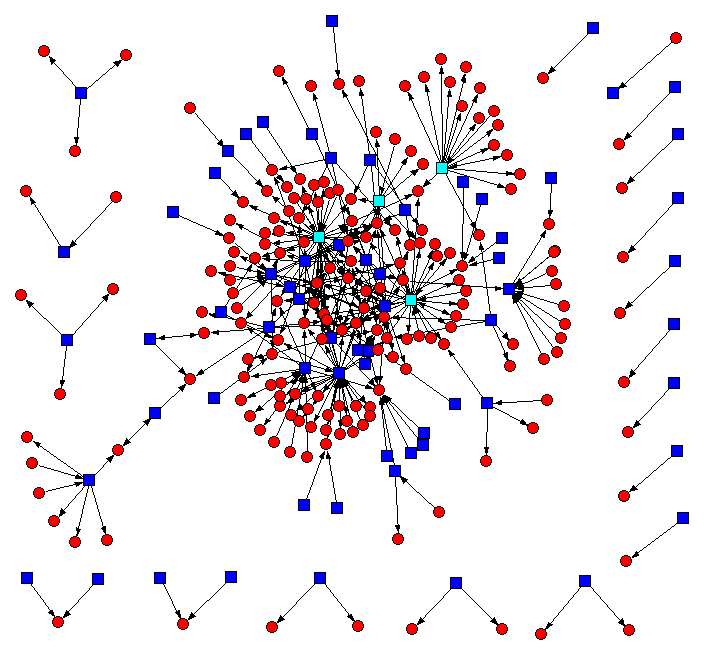}}
    \caption{\csentence{FDR networks for 1984 and 2009}
Statistically validated FDR networks for the calendar years 1984 (left) and 2009 (right). Banks are shown as squares and firms as red circles. Banks labeled by the Cyan color are City Banks whereas all the other banks are labeled as Blue.  Arcs are starting from the node performing the statistical validation. A number of bidirectional arcs are observed (25 in 1984 and 9 in 2009).}
      \label{FDRexamples}      
      \end{figure}

In Fig. \ref{FDRexamples} we show the FDR networks of 1984 and 2009 (the first is the one with the highest fraction of elements in the LCC whereas the second is the one with the lowest fraction). The two FDR networks summarize the evolution observed both in the original credit network and in the filtered credit network. Specifically the networks evolve from highly interconnected networks to sparser networks where the dependence of individual banks form firms and viceversa is much more simplified. This type of representation of the credit linkages suggests that the Japanese credit market has been subjected to a process of simplification starting from the year 2000 (see the temporal evolution of the number of edges in Table \ref{Table1}). Although the majority of elements are still part of a large connected component in the FDR network the interlinkages among banks and firms are getting simpler as time elapses. 

\begin{figure}[h!]
  	 {\includegraphics[scale=0.41]{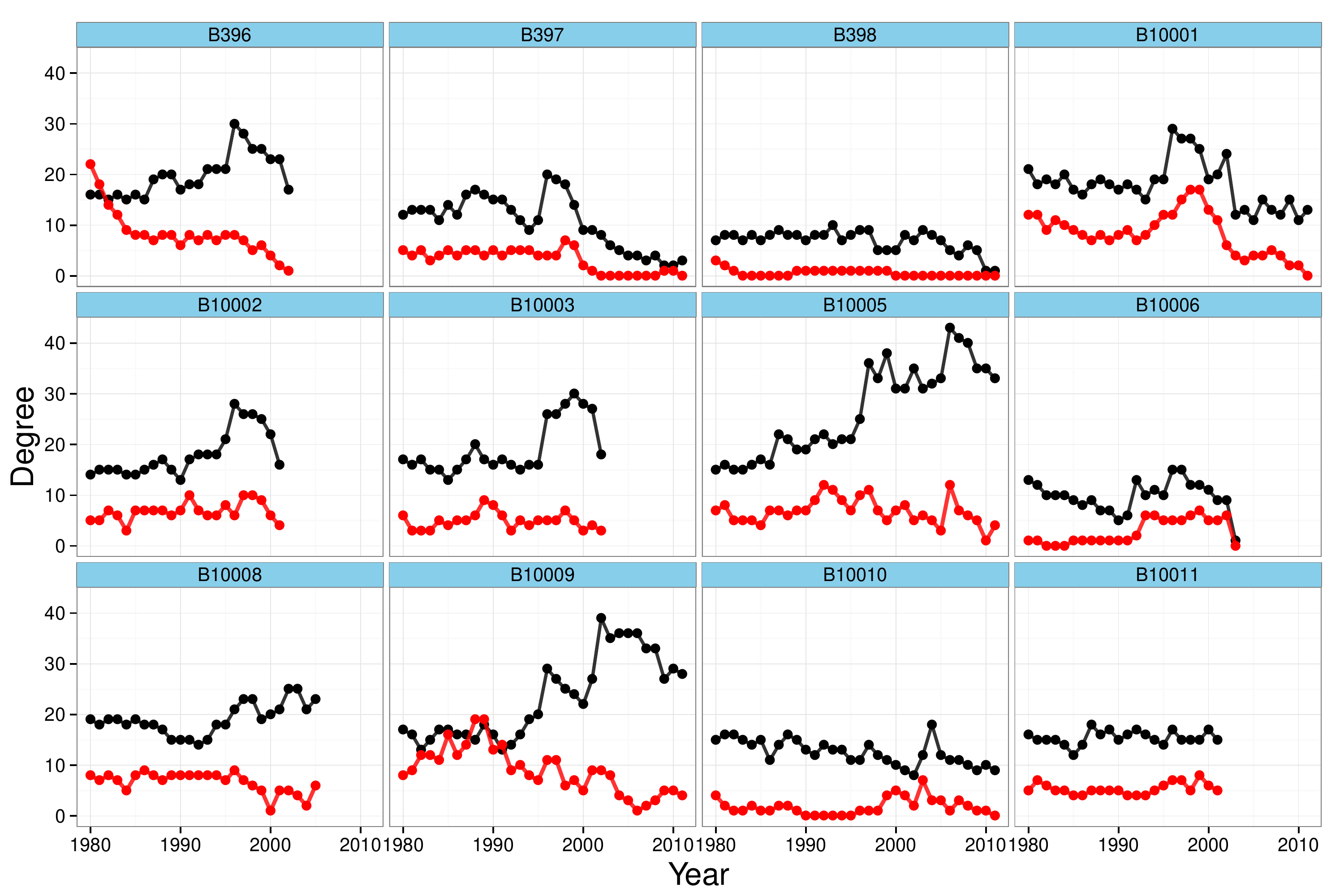}}
    \caption{\csentence{Degree evolution of Long term banks and City Banks}
Time evolution of the in-degree (red symbols) and out-dregee (black symbols) of 3 Long term banks (B396 Industrial Bank of Japan, B397 Long Term Bank of Japan, and B398 Nippon Credit Bank) and 9 City banks (B10001 Mizuho Bank in 2011, B10002 Sakura Bank in 2001, B10003 Fuji Bank in 2002, B10005 Bank of Tokyo - Mitsubishi UFJ in 2011, B10006 Asahi Bank in 2003, B10008 UFJ Bank in 2006, B10009 Sumitomo Mitsui Banking Corp. in 2011, B10010 Resona Bank in 2011, and B10011 Tokai Bank in 2002. The degree refer to the FDR statistically validated networks. The time evolution is halted when the bank undergoes a merger with another bank. For details about the merging process see the main text.}
\label{DegreeEv}      
\end{figure}

In the Japanese credit market the so called City banks play an important role. It is therefore of interest to track in detail the time evolution of the in-degree and out-degree of these banks in the statistically validated FDR networks. This information is shown in Fig. \ref{DegreeEv} where we report the time evolution of 3 Long-term banks and 9 City banks. During the years some City banks have undergone 
merging with other City banks or other financial institutions. This is the reason why some of the City banks disappear at a given year. 

By analyzing Fig. \ref{DegreeEv} we note that the out-degree (black symbols) is almost always higher than the in-degree (red symbols). This means that for each City bank the number of firms which are receiving statistically validated loans from the bank, i.e. the loans that are highly relevant for the bank, is higher than the number firms that have their loans highly dependent on the bank,  i.e. loans that are highly relevant for the firm. We also note that the merging and acquisition processes make some of the City banks (for example B10005, B10009)  highly exposed to many different firms. This exposure is showing that the process of full diversification of risk is only partially achieved. 

\begin{figure}[h!]
  {\includegraphics[scale=0.23]{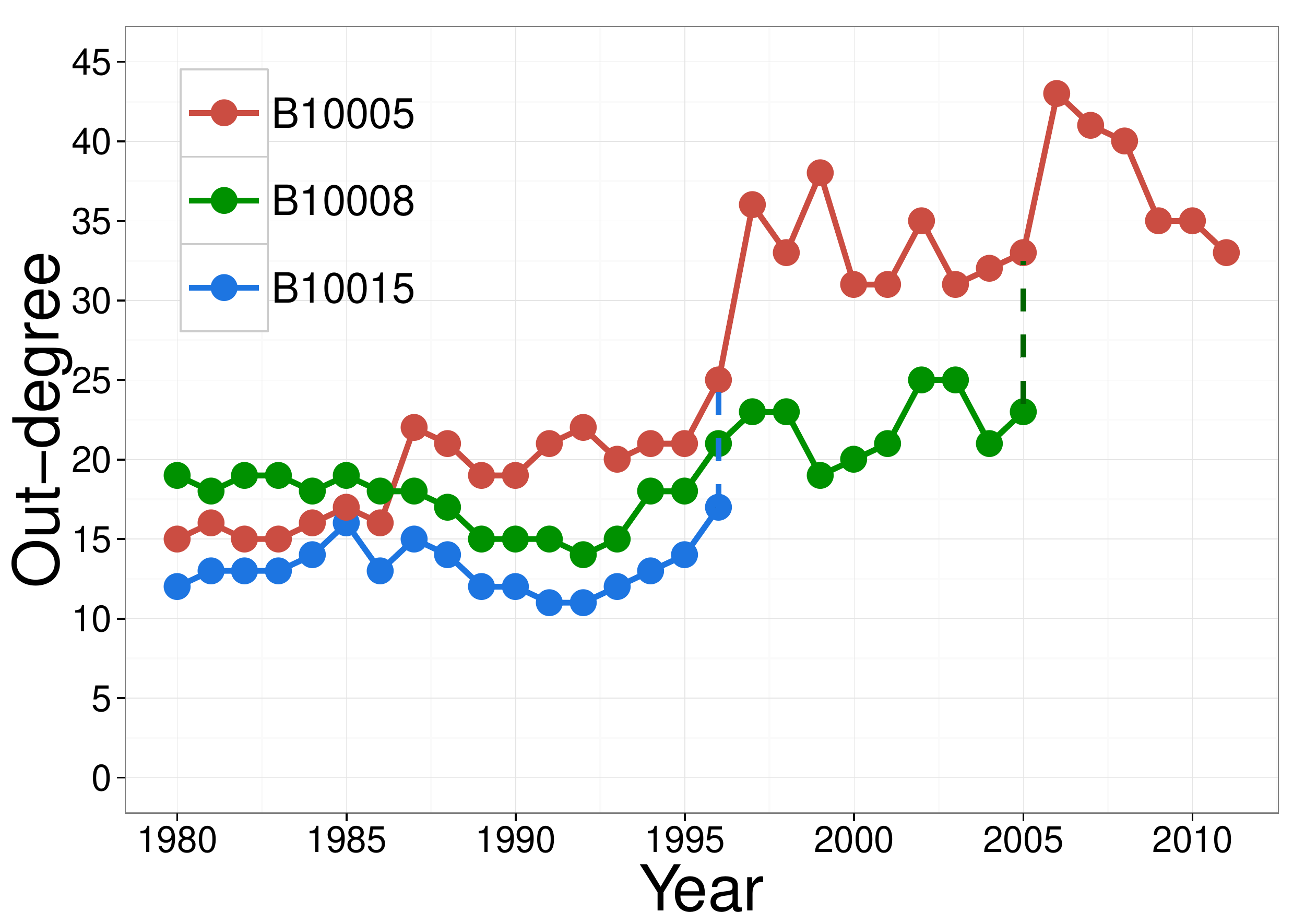}
  \includegraphics[scale=0.23]{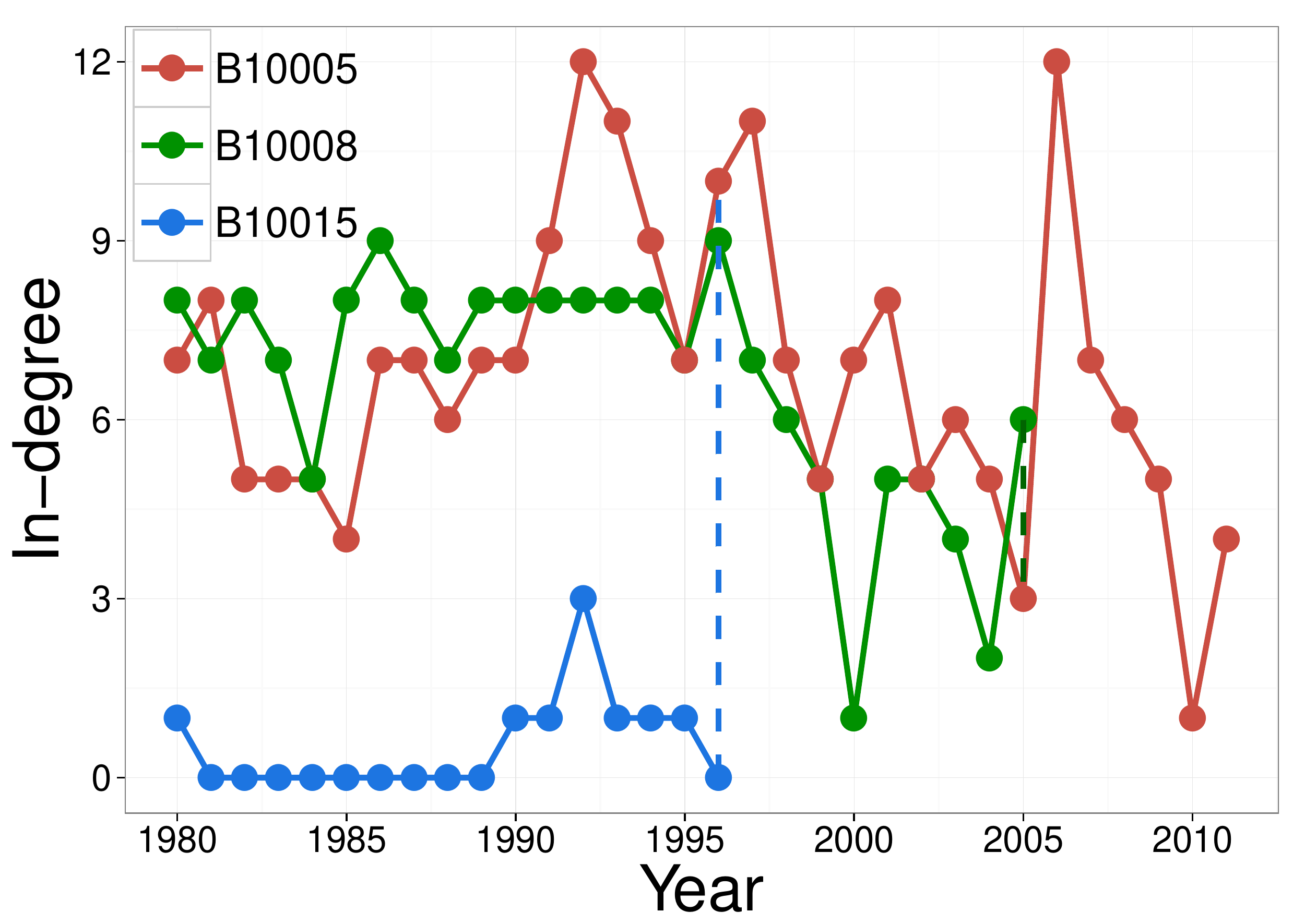}}
  
    \caption{\csentence{Out-degree and in-degree of three city banks.}
Out-degree (left panel) and in-degree (right panels) of banks labeled as B10005, B10008, and B10015 in the database. B10008 was the bank the Sanwa Bank from 1980  to 2002 and become the UFJ Bank Ltd in 2006. B10015 was the Bank of Tokyo from 1980 to 1996 when it merged with Mitsubishi Bank. B10005 code refers to Mitsubishi Bank from 1980 to 1996, to Bank of Tokyo - Mitsubishi Bank from 1996 to 2002 and to Bank of Tokyo - Mitsubishi - UFJ bank from 2002 to 2011.}
      \label{BTMU}      
      \end{figure}

To clarify how the database deals with merging and acquisition and to present a case study of evidence of the increasing number of our-degree in the FDR network we track the process of merging and acquisition of the Banks B10005, B10008, and B10015. In 1980 these codes of the database were associated with the banks Mitsubishi Bank, UFJ Bank Ltd., and Bank of Tokyo respectively. In 1996 the banks B10005 (at that time Mitsubishi Bank) and B10015 (Bank of Tokyo) merged to form the bank Bank of Tokyo-Mitsubishi that since 1997 used the code B10005. In 2006 another merging occurred. Specifically, B10005 (Bank of Tokyo-Mitsubishi) and B10008 (UFJ Bank) merged into the bank (Bank of Tokyo-Mitsubishi UFJ) that since 2007 continued to use the database code B10005. The process of merging and time evolution of the in-degree and out-degree of these banks are shown in Fig. \ref{BTMU}. From the figure we note that immediately after the merging the out-degree of the new bank is always higher that the out-degree of the two banks that merged. In other words the data show the need of some time to improve the diversification of the loans of the new bank. In fact after the merging a gradual decrease of the out-degree is observed. The time evolution of the in-degree show the relevance of the loans of the new bank for the firms. The right panel of Fig. \ref{BTMU} shows that the merging of 1996 of Bank of Tokio with Mitsubishi Bank had no big impact on relevance of loans for the  firms having credit relationships with them whereas the merging of 2006 implied that a certain number of firms had to consider their loans with the new bank as more relevant for them then in the period before merging.

 \begin{table}[h!]
 \centering
   \caption{Summary for over-expressions of attributes in statistically validated FDR networks. We indicate different economic sectors with a one or two letter code as follows: C Construction, CL Credit \& Leasing, FMP Fish \& Marine 	
   	Products, P Petroleum, RT Railroad Transportation, RE Real Estate, ST Sea Transportation, SH Security Houses, 
    UE Utilities Electric, UG Utilities Gas, CB city banks, LI life-insurance banks.}
	\label{Table2}
	 \resizebox{\columnwidth}{!}{%
      \begin{tabular}{l|cccccccccc|c|cc}
        \hline
          Year & C & CL & FMP &  P & RT &%
           RE &  ST. & SH & UE & UG & Tokyo & CB & LI\\           
        \hline
        1980 & -- & -- & OE & OE & -- & -- & OE & -- & OE & -- & OE & OE & --\\
        1981 & -- & -- & -- & OE & -- & -- & OE & -- & OE & -- & OE & OE & --\\
        1982 & OE & -- & -- & OE & -- & -- & -- & -- & OE & -- & OE & OE & --\\
        1983 & -- & -- & -- & OE & -- & -- & -- & -- & OE & -- & OE & OE & --\\
        1984 & -- & -- & -- & OE & -- & -- & OE & -- & OE & -- & OE & OE & OE\\
        1985 & -- & -- & -- & OE & OE & -- & -- & -- & OE & -- & OE & OE & OE\\
        1986 & -- & -- & -- & OE & OE & -- & OE & -- & OE & -- & -- & OE & OE \\
        1987 & -- & OE & -- & OE & -- & -- & OE & -- & OE & -- & OE & -- & --\\
        1988 & -- & OE & -- & -- & -- & -- & OE & -- & OE & -- & OE & -- & --\\
        1989 & -- & OE & -- & OE & -- & OE & OE & -- & OE & -- & OE & -- & --\\
        1990 & -- & OE & -- & -- & OE & OE & OE & -- & OE & -- & OE & -- & --\\
        1991 & -- & OE & -- & -- & -- & OE & OE & -- & OE & -- & OE & -- & --\\
        1992 & -- & OE & -- & -- & -- & OE & OE & -- & OE & -- & OE & -- & --\\
        1993 & OE & OE & -- & -- & -- & OE & OE & -- & OE & -- & OE & -- & --\\
        1994 & OE & OE & -- & -- & OE & OE & -- & -- & OE & OE & OE & -- & --\\
        1995 & OE & OE & -- & -- & OE & OE & -- & -- & OE & -- & OE & -- & --\\
        1996 & OE & OE & -- & -- & OE & OE & -- & OE & OE & -- & OE & -- & --\\
        1997 & OE & OE & -- & -- & OE & OE & -- & OE & OE & -- & OE & -- & --\\
        1998 & OE & OE & -- & -- & OE & OE & -- & -- & OE & -- & OE & -- & --\\
        1999 & OE & OE & -- & -- & OE & OE & -- & -- & OE & -- & OE & -- & --\\
        2000 & -- & OE & -- & -- & OE & OE & -- & -- & OE & -- & -- & -- & --\\
        2001 & OE & OE & -- & -- & OE & OE & -- & -- & OE & -- & -- & -- & --\\
        2002 & -- & OE & -- & -- & OE & OE & -- & -- & OE & -- & -- & -- & --\\
        2003 & -- & OE & -- & -- & OE & OE & -- & -- & OE & -- & OE & -- & --\\
        2004 & -- & OE & -- & -- & OE & -- & -- & -- & OE & -- & OE & -- & --\\
        2005 & -- & OE & -- & -- & OE & -- & -- & OE & OE & -- & OE & -- & --\\
        2006 & -- & OE & -- & -- & OE & -- & -- & -- & OE & -- & OE & -- & --\\
        2007 & -- & OE & -- & -- & OE & OE & -- & -- & OE & -- & OE & -- & --\\
        2008 & -- & OE & -- & -- & OE & OE & -- & -- & OE & -- & OE & -- & --\\
        2009 & -- & OE & -- & -- & OE & -- & -- & -- & OE & -- & -- & -- & --\\
        2010 & -- & OE & -- & -- & OE & -- & -- & -- & OE & -- & -- & -- & --\\
        2011 & -- & OE & -- & -- & OE & -- & -- & -- & OE & -- & -- & -- & --\\
        \hline
      \end{tabular}
      }
\end{table}

 \subsection*{Comparison of original, filtered and reciprocal credit network}
With our procedure we therefore have obtained three networks. The first network is the original undirected bipartite network, the second network is a filtered directed network showing the backbone of the credit relationships selected by estimating the loans that are over-expressed for each bank and/or each firm when the testing is performed against a null hypothesis assuming equal diversification of the loans provided to firms (when the focus is on banks) or of the loans obtained by the firm (when the focus is on the firms). The third network is a subgraph of the second one containing only nodes connected by reciprocal links.
A summary statistics of the nodes and links of these networks is given in Table \ref{Table1}. Here we report on the attributes characterizing filtered and the reciprocal networks. We use as attributes the type of a bank, the economic sector of a firm and the geographical location of a firm (defined as the prefecture where the firm is registered). The over-expression of the attributes is estimated by using the procedure illustrate in reference \cite{Tumminello2011}. This is a urn type inspired statistical test performing multiple hypothesis comparison. The multiple hypothesis test correction in this case is a Bonferroni correction \cite{Miller1981}.

\begin{figure}[h!]
{\includegraphics[scale=0.42]{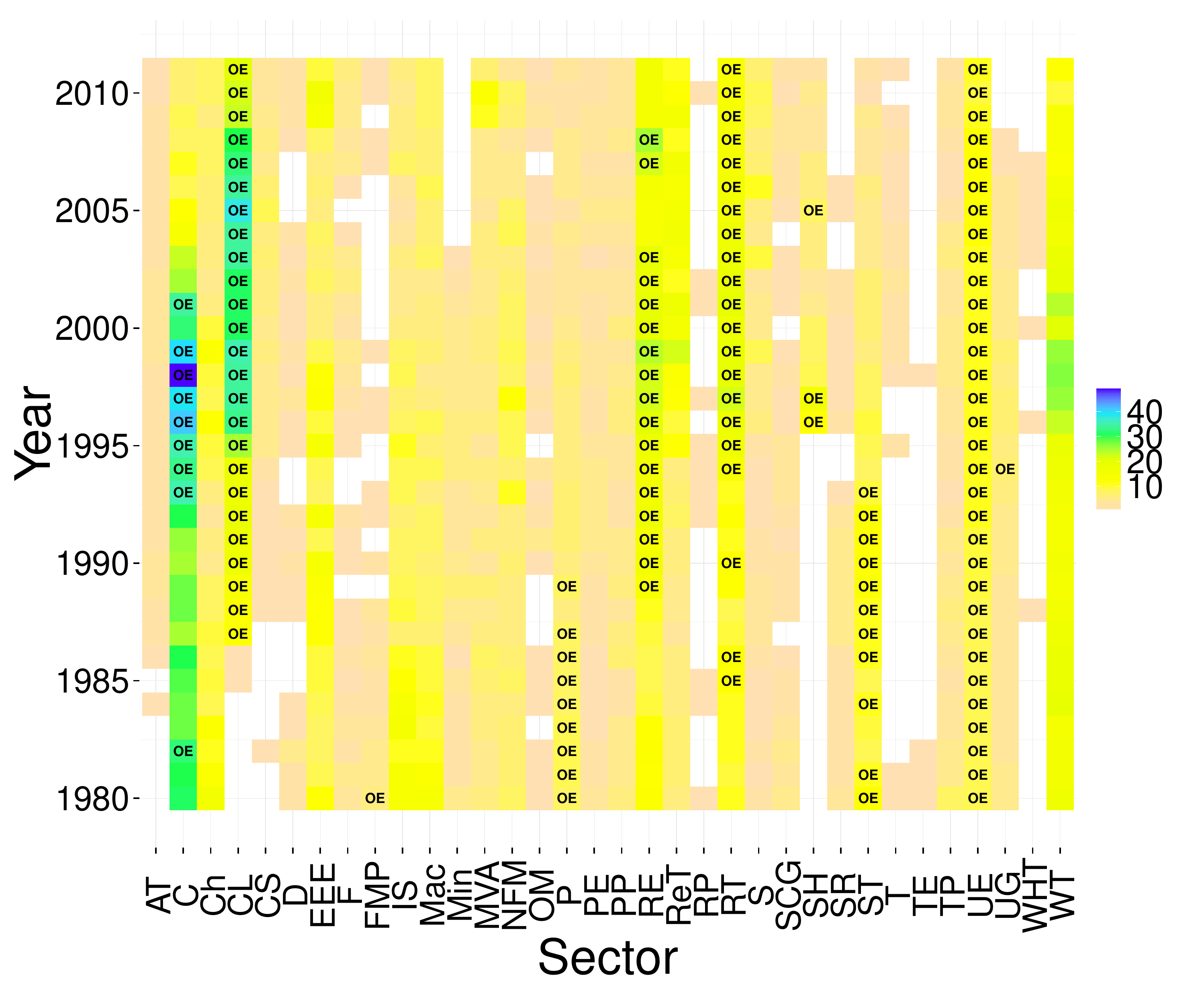}}
      \caption{\csentence{Color code summary of the number of firms in a specific economic sector observed in the statistically validated FDR graph}. The label OE (over-expressed) indicates the statistical validation of the attribute for the corresponding year. The codes used 
      for sectors on the x axis can be read as follows: \textbf{AT} Air Transportation, 
      \textbf{C} Construction, \textbf{Ch} Chemicals, \textbf{CL} Credit \& Leasing, 
      \textbf{CS} Communication Services, \textbf{D}  Drugs, \textbf{EEE} Electric \& Electronic Equipment, 
      \textbf{F} Foods, \textbf{FMP} Fish \& Marine Products, 
      \textbf{IS} Iron \& Steel, \textbf{Mac} Machinery, 
      \textbf{Min} Mining, \textbf{MVA} Motor Vehicles \& Auto Parts, \textbf{NFM} Non Ferrous Metal \& Metal Products, 
      \textbf{OM}  Other Manufacturing,  \textbf{P} Petroleum, \textbf{PE} Precision Equipment, 
      \textbf{PP} Pulp \& Paper, \textbf{RE} Real Estate, \textbf{ReT} Retail Trade, 
      \textbf{RP} Rubber Products, \textbf{RT} Railroad Transportation, \textbf{S} Services, \textbf{SCG} Stone, Clay \& Glass Products, 
      \textbf{SH} Securities Houses, \textbf{SR} Shipbuilding \& Repair, \textbf{ST} Sea Transportation, \textbf{T} Trucking, 
      \textbf{TE} Transportation Equipment, \textbf{TP} Textile Products, 
      \textbf{UE} Utilities - Electric, \textbf{UG} Utilities - Gas, \textbf{WHT} Warehousing \& Harbor Transportation , 
      \textbf{WT} Wholesale Trade.}
      \label{OE_sec}      
      \end{figure}

The results of the test for the filtered graph are summarized in Table \ref{Table2}. The Table shows that the loans with the highest concentration of money involves primarily City banks and Life insurance banks, firms whose headquarter is located in the prefecture of Tokyo, and firms of  Construction, Credit \& Leasing, Fish \& Marine Products, Petroleum, Railroad Transportation, Real Estate, Sea Transportation, Security Houses, Utilities Electric, and Utilities Gas economic sectors. The most over-expressed economic sectors are certainly  Credit \& Leasing and  Utilities Electric economic sectors. In the analysis of Table \ref{Table2}, it is worth noting that over-expression is not necessarily indicating a large frequency of the observed attribute in the filtered network. In fact the test is detecting {\it deviation} from the null hypothesis assuming heterogeneity of the frequency of the different attributes. To make this point clear in Fig. \ref{OE_sec} we show a color code representation of the number of economic sector attributes detected in the filtered network together with the information whether each specific attribute is over-expressed or not with respect to the heterogeneity observed in the original network. From Fig. \ref{OE_sec} it can be seen that the sectors  of Construction and Credit \& Leasing  are characterized by both a high value of the number of attributes and over-expression. However a high value of attributes does not guarantee an over-expression of the attributes as it is seen for the case of firms in Wholesale Trade. On the contrary, even when the value of the number of attributes is low we can detect over-expression if the firms turn out to be highly present in the filtered network as it is the case for the Utilities Electric economic sector. Intermediate cases are observed for the Railroad Transportation and Real Estate economic sectors.

From Fig. \ref{OE_sec} and Table \ref{Table2} we notice that the mostly over-expressed sectors are the sectors of  Construction, Credit \& Leasing, Railroad Transportation, Real Estate and Utilities Electric. Most of these sectors do not overlap with the over-expressed economic sectors observed in the bipartite clusters when one performs a community detection on the unweighted bipartite graph \cite{Marotta2015}. It should be noted that a priori there is no reason to expect an overlap of the over-expression because the community detection on the unweighted bipartite network is providing information about the mesoscopic organization of the credit network whereas the present information is filtering the role of the most intense credit relationships from the perspective of banks and/or firms.           

In our analysis the credit relationships that are validated both for the lending bank and for the borrowing firms are of special interest because they are highlighting a strong interdependence between the bank and the firm. The summary statistics of Table \ref{Table1} shows that the number of these strong interdependences has been decreasing in Japan starting from 1998 suggesting a improvement of the diversification processes of credit allocation in the Japanese market.

\begin{figure}[h!]
  {\includegraphics[scale=0.2]{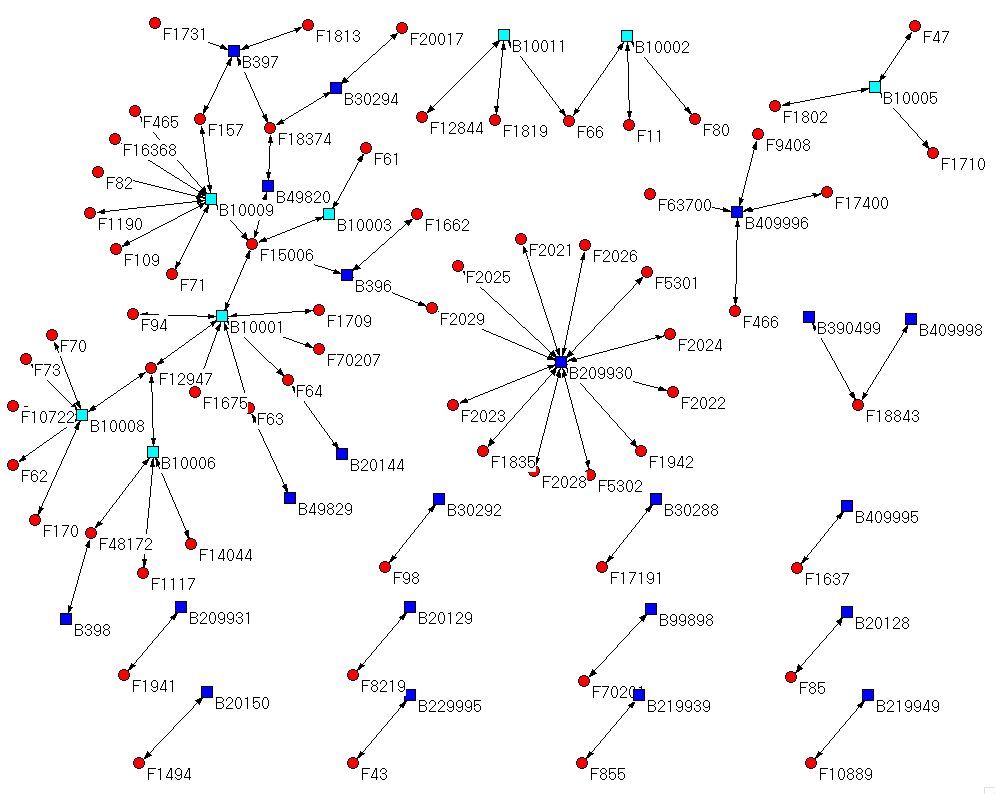}
  \includegraphics[scale=0.19]{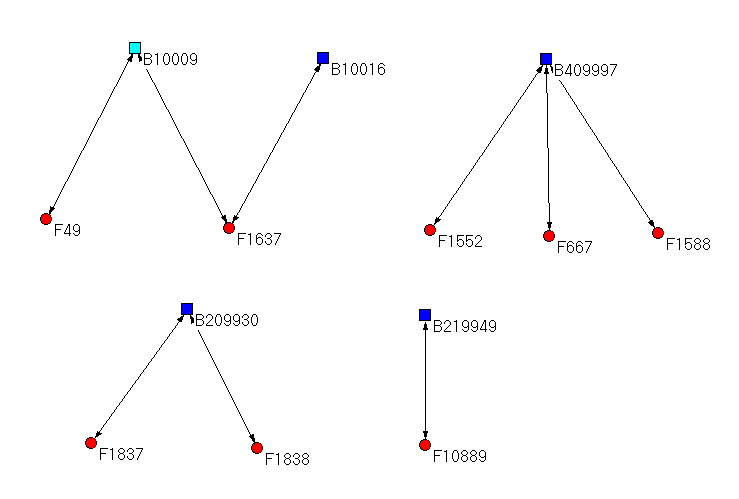}}
    \caption{\csentence{Networks of bidirectional validated links}
Statistically validated links of the FDR networks that are bidirectional for the calendar years 1997 (left) and 2011 (right). Banks are shown as squares and firms as red circles. Bank symbols with the Cyan color are City Banks whereas all the other banks are shown as Blue squares. The year 1997 is the year with the maximal number of bidirectional validated links whereas 2011 is the year with the minimal number.}
      \label{BidirectionalEX}      
      \end{figure}

In Fig. \ref{BidirectionalEX} we show two examples of the bidirectional links that are present in the statistically validated FDR networks. The first example (left panel) refers to the year 1997 which is the one characterized by the highest number of bidirectional links. The subnetwork contains 11 City banks (Cyan boxes) and moreover 8 of them are present in the largest connected component comprising 12 banks and 30 firms. In other words a strong interdependence of the credit relationships involving the majority of the City banks and a large number of firms was present in that year. Two other City banks are strongly interconnected with 6 firms. The subnetwork of bidirectional links in 2011 is completely different. In this case only two City banks are present and the highlighted credit relations are only with two firms.

\begin{figure}[h!]
{\includegraphics[scale=0.46]{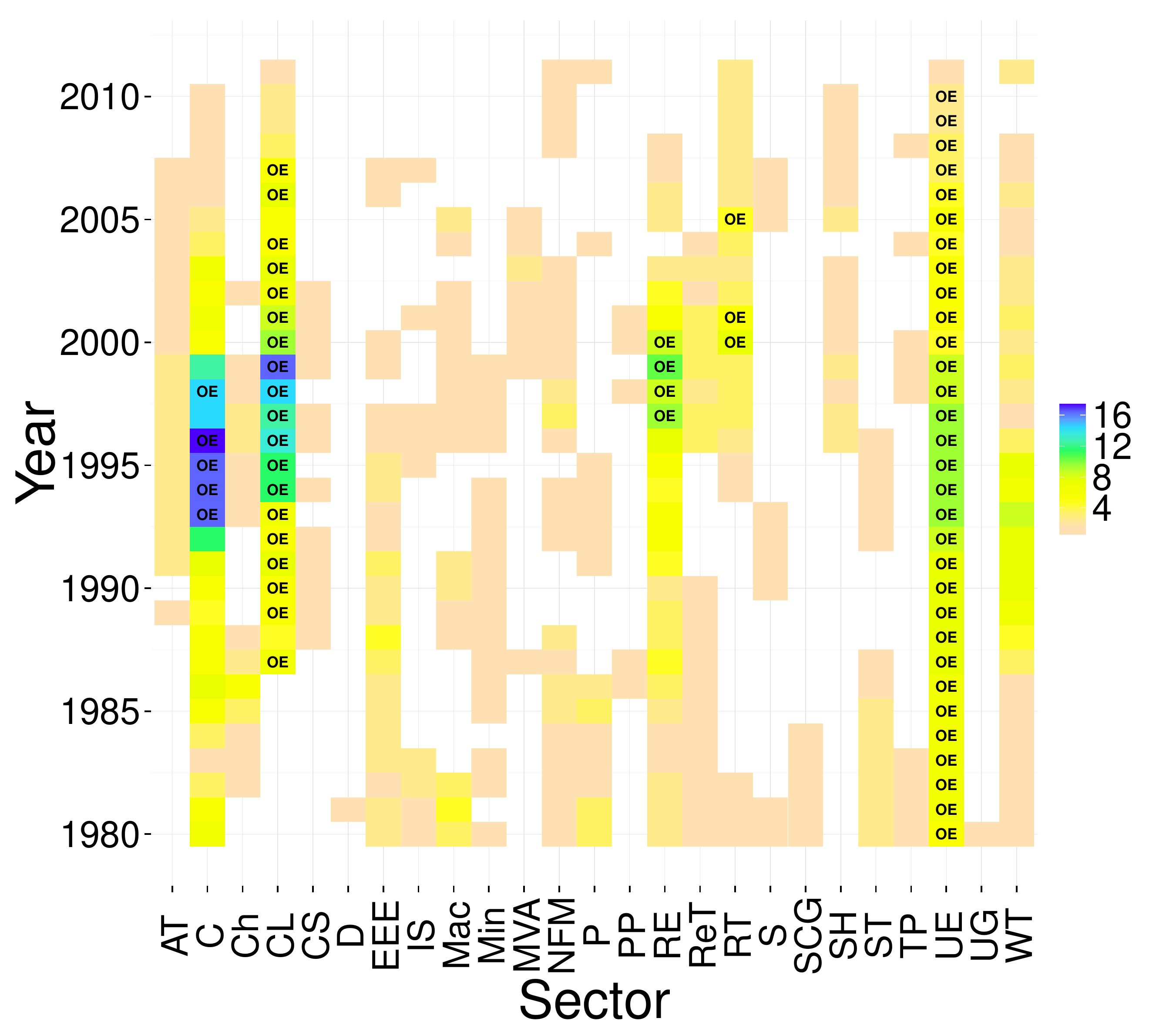}}
      \caption{\csentence{Color code summary of the number of firms of economic sector observed in the subgraphs of the FDR validated networks with bidirectional links}. The label OE (over-expressed) indicates the statistical validation of the attribute for the corresponding year.
      The label OE (over-expressed) indicates the statistical validation of the attribute for the corresponding year. The codes used 
      for sectors on the x axis can be read as follows: \textbf{AT} Air Transportation, 
      \textbf{C} Construction, \textbf{Ch} Chemicals, \textbf{CL} Credit \& Leasing, 
      \textbf{CS} Communication Services, \textbf{D}  Drugs, \textbf{EEE} Electric \& Electronic Equipment, 
      \textbf{IS} Iron \& Steel, \textbf{Mac} Machinery, 
      \textbf{Min} Mining, \textbf{MVA} Motor Vehicles \& Auto Parts, \textbf{NFM} Non Ferrous Metal \& Metal Products, 
	  \textbf{P} Petroleum, \textbf{PP} Pulp \& Paper, \textbf{RE} Real Estate, \textbf{ReT} Retail Trade, 
      \textbf{RT} Railroad Transportation, \textbf{S} Services, \textbf{SCG} Stone, Clay \& Glass Products, 
      \textbf{SH} Securities Houses, \textbf{ST} Sea Transportation, \textbf{TP} Textile Products, 
      \textbf{UE} Utilities - Electric, \textbf{UG} Utilities - Gas, \textbf{WT} Wholesale Trade.}
      \label{OE_Mut_sec}      
      \end{figure}

We show information about the type of economic sectors present in the subnetwork of bidirectional links and about the over-expression of some of them in Fig. \ref{OE_Mut_sec}. The figure shows that the economic sectors with the largest number of firms are Construction, Credit \& Leasing, Real Estate, Utilities Electric, and Wholesale Trade. All these sectors with the exception of Wholesale trades present also over-expression of presence with respect to the frequency of the attributes present in the original networks. In others words firms of these economic sectors have been the sours of the major degree of interlinkages in the Japanese credit market. It is also worth noting that the time period close to 1997 was the period of maximal values of number of firms and over-expressions of the cited economic sectors.
      
\begin{figure}[h!]
{\includegraphics[scale=0.5]{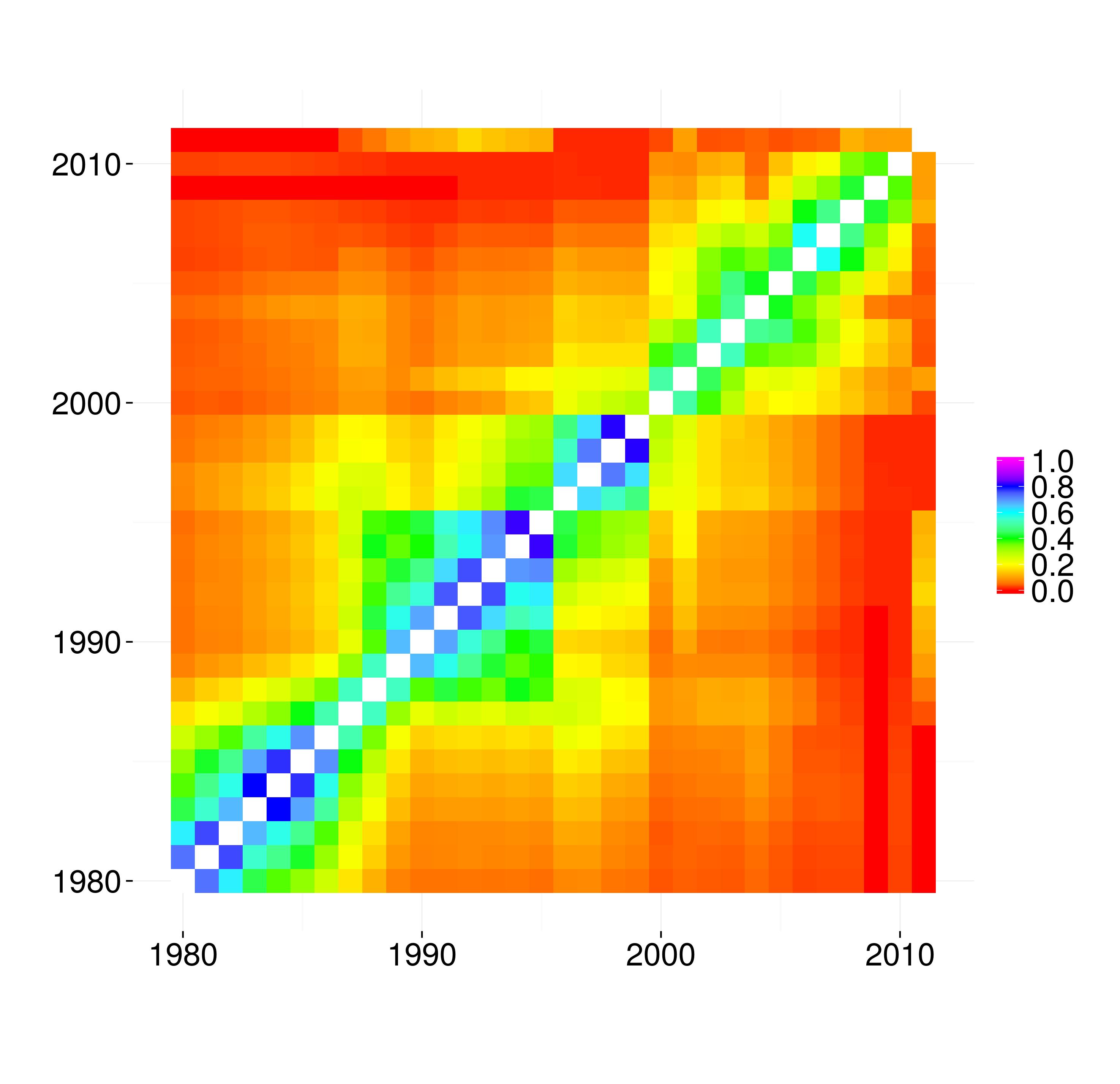}}
      \caption{\csentence{Color code summary of the weighted Jaccard measure computed between all pairs of the sub-networks of bidirectional validated FDR links obtained for the different calendar years.}}
      \label{JaccM}      
      \end{figure}

However the degree of interlinkage has changed over time. To quantitatively evaluate the change of the subgraph of the FDR network with bidirectional links we evaluate the weighted Jaccard measure between all pairs of subgraphs obtained for the different calendar years. For the definition of the weighted Jaccard measure in a similar type of network time evolution investigation see ref. \cite{Hatzopoulos2015,Iori2015}. The results of this estimation is shown in Fig. \ref{JaccM} where we summarize the values obtained for all pairs of years by using a matrix where the value of each element is shown according to a color code. The figure shows that the sub-networks were relatively stable during the periods 1980-1987, 1988-1995, and 1994-1999. Starting from 2000 the degree of stability has significantly decreased in parallel with the shrinking of the sub-network.

\section*{Conclusions}
In this paper we investigate the Japanese credit market by using a database of the credit loans provided by Japanese banks to large Japanese firms at a micro level, i.e. we have information about the single bank-firm credit relationship. The database covers a long period of time (1980-2011) that has seen large crises of the Japanese banking sector and a sequence of merging of largest banks.

By adapting a filtering method of weighted networks proposed in ref. \cite{Serrano2009} we select those credit relationships that are not compatible with a null hypothesis assuming uniform distribution of the loans. The test was performed for each credit relationship both from the perspective of a bank and from the perspective of the firm. We have used a multiple hypothesis test correction needed to avoid false positive in the statistical validation of networks \cite{Tumminello2011b}. Specifically we have used the false discovery rate correction.

Our results show the existence of a backbone of the credit relationships not compatible with the null hypothesis of uniform diversification of the bank or of the firm. The nature of this backbone has changed over time both in size, fraction of credit and attributes characterizing the firms. Of major interest we believe are the bidirectional links observed in the FDR statistically validated networks. This bidirectional links are indicating that the credit relation should be of great importance both for the bank and for the firm. This observation makes these links of special importance as channels of potential dependency of a firm toward a bank and viceversa. In the investigated period, it is worth noting that the number and strength of these links has first increased from 1980 to 1997 and then it has decreased since then. The 1997 year is a very special year for the banking system of Japan because this was the year of the largest recent systemic crisis of the Japanese banking system \cite{Kanaya2000}. Our analysis of the most pronounced interlinkages show that the changes introduced in 1997 and also in 2000 and the series of merging of the major banks have significantly reduced in the recent years the level of special interlinkages present between banks and firms in the Japanese credit system.


\begin{backmatter}

\section*{Competing interests}
  The authors declare that they have no competing interests.

\end{backmatter}
\end{document}